\documentclass[preprint,prc,aps,nofootinbib,superscriptaddress]{revtex4}
\usepackage{graphicx}
\usepackage{epsfig}
\usepackage{bm}
\usepackage{amssymb}

\begin{document}

\title{Magnetic moment of hyperons in nuclear matter by using quark-meson coupling models}

\author{C. Y. Ryu} \email{cyryu@color.skku.ac.kr}
\affiliation{Department of Physics and Institute of Basic Science,
Sungkyunkwan University, Suwon 440-746, Korea}

\author{C. H. Hyun} 
\affiliation{Department of Physics Education, Daegu University,
Gyeongsan 712-714, Korea}

\author{T.-S. Park} 
\affiliation{Department of Physics and Institute of Basic Science,
Sungkyunkwan University, Suwon 440-746, Korea}

\author{S. W. Hong} 
\affiliation{Department of Physics and Institute of Basic Science,
Sungkyunkwan University, Suwon 440-746, Korea}

\date{July 25, 2008}

\begin{abstract}
We calculate the magnetic moments of hyperons
in dense nuclear matter by using relativistic quark models.
Hyperons are treated as MIT bags, and the interactions are 
considered to be mediated by the exchange of 
scalar and vector mesons which are approximated as mean fields.
Model dependence is investigated by using the quark-meson coupling model
and the modified quark-meson coupling model; in the former the
bag constant is independent of density and in the latter it depends on density.
Both models give us the magnitudes of the magnetic moments 
increasing with density for most octet baryons. 
But there is a considerable model dependence in the values of the magnetic
moments in dense medium. The magnetic moments
at the nuclear saturation density calculated by the quark meson coupling model
are only a few percents larger than those in free space, 
but the magnetic moments from the modified quark meson coupling model
increase more than 10~\% for most hyperons.
The correlations between the bag radius of hyperons
and the magnetic moments of hyperons in dense matter are discussed.
\end{abstract}


\maketitle


\section{Introduction}

Deep inelastic muon-nucleus scattering in the European
Muon Collaboration showed that the electromagnetic (EM)
properties of the nucleon in nuclear medium could be different
from those in free space \cite{emc83}.
It was shown that the magnetic moment of the proton in $^{12}{\rm C}$ seemed enhanced
by about 25~\% compared to that in free space \cite{mulders90}.
Experiments were performed to explore various observables that could indicate
medium effects on the EM properties of the nucleon, such as longitudinal
response function, polarization transfer, induced polarization, and etc.
A very recent experiment at JLab \cite{jlab07} provides another
positive indication of the medium modification of 
the EM form factors of the nucleon.
On the theoretical side, several models that treat the nucleon
as a composite system of quarks were proposed to calculate
the in-medium EM form factors of the nucleon
\cite{cheon92,frank96,lu99,yak03,smith04,horikawa05}.
A cloudy bag model calculation \cite{cheon92} predicted a substantial
enhancement of the magnetic moment in the nuclear medium
in the range 2--20~\%.
On the other hand, models such as light-front constituent
quark model \cite{frank96}, quark-meson coupling model with pion 
cloud \cite{lu99}, Skyrme model \cite{yak03}, chiral quark
soliton model \cite{smith04} and Nambu-Jona-Lasino model \cite{horikawa05}
give less enhancement, up to 10~\% at most.
Quasi-elastic electron-nucleus scattering was expected to provide
possible indications for the in-medium modification of nucleon structure
functions, and investigations along this line were performed.
In Refs.~\cite{kc03,kw03}, the cross sections for
quasi-elastic $(e,\, e')$ scattering were calculated
with and without the in-medium EM form factors.
But the results showed no clear indication of medium modification.
Although much work has been done in both theory and experiment, 
the situation still remains controversial,
especially for the magnetic moment.

In this work we calculate the magnetic moments of the
octet baryons in nuclear matter with the quark-meson coupling (QMC)
\cite{qmc88} and the modified quark-meson coupling (MQMC) \cite{mqmc96}
models.
Based on the MIT bag model, these models provide a simple but robust tool
for the description of baryon properties in free space and bulk properties
of symmetric and asymmetric nuclear matter \cite{ST-PLB}
and neutron stars \cite{MPP,RHHJ}. 
Saito and Thomas calculated the in-medium magnetic moment of the
proton in symmetric matter with the QMC model \cite{st95}.
An interesting result in their work is the dependence 
of the magnetic moment on the bag radius.
Three values of the bag radius, 0.6, 0.8 and 1.0 fm were adopted for
the proton in free space.
Changes in the values of the magnetic moments due to the medium from those     
in free space are relatively small, but
the changes depend considerably on the bag radius in free space.
For instance, the magnetic moment in medium at the nuclear saturation density
is only about 1~\% larger than that in free space if the bag radius is 
chosen as 0.6 fm, but if the bag radius is 1.0 fm 
the magnetic moment becomes about 7~\% larger.

We shall investigate the model dependence of the magnetic moments in medium
by considering both the QMC and MQMC models.
While the QMC model has the problem of yielding too small a spin-orbit
interaction, the MQMC model with the density-dependent bag constant
produces the magnitudes of $\sigma$ and $\omega$ meson fields
similar to those obtained from the Dirac phenomenology and 
quantum hadrodynamics, which produce the right magnitudes of the spin-orbit 
interaction. 
The MQMC model is also able to reproduce the nuclear
saturation properties better than the QMC model,
which will be presented in Sec.IIA.
A big difference between the bag properties obtained from QMC and MQMC
is in the behavior of the bag radius in nuclear matter.
In the QMC model, the bag radius decreases as density increases,
but in the MQMC model the bag radius increases with density \cite{JJ}.
Since the magnetic moment depends on the bag radius,
it is expected that the prediction of the magnetic moment from 
the MQMC model will differ from that obtained from the QMC model.

The magnetic moment of a hyperon in medium was experimentally 
studied only recently. The magnetic moment of a $\Lambda$ hyperon
in a hypernucleus $^{7}_{\Lambda}{\rm Li}$ has been measured at BNL
\cite{tamura07}.
The result is still preliminary with large errors.
Further experiments are needed, for example, in J-PARC \cite{tamura07}
to determine the in-medium EM properties of the hyperon.
Thus it is timely to investigate the magnetic moment of hyperons
in medium theoretically.

In Sec. II, basic ingredients of the models are presented.
The magnetic moments of octet baryons are expressed in terms of the quark
wave functions.
Sec. III shows numerical results, and discussions on 
the model dependence and the correlation between the bag radius
and the magnetic moments follow.
Sec. IV summarizes the paper.

\section{Models}
\subsection{QMC and MQMC models for
nuclear matter \label{sec:matter}}

In the QMC model a nucleon in nuclear matter
is described by a static MIT bag in which
quarks couple to meson fields that are treated as mean fields.
The quark field $\psi_q$ inside the bag satisfies the Dirac equation
\begin{eqnarray}
\left[ i \gamma \cdot \partial - ( m_q - g^q_\sigma\, \sigma)
- g^q_\omega\, \gamma^0 \, \omega_0 \right]\, \psi_q = 0,
\end{eqnarray}
where $m_q$ $\left( q= u,\, d,\, s\right)$ is the bare quark mass,
$\sigma$ and $\omega_0$ are the mean fields of $\sigma$ and
$\omega$ mesons, respectively,
and
$g^q_\sigma$ and $g^q_\omega$ are the quark-meson coupling constants.
Here, we assume $m_u=m_d=0$ and $m_s = 150$ MeV.

The ground state solution of the Dirac equation is given by
\begin{eqnarray}
\psi_q(\bm{r},\, t) =
{\cal N}_q \exp(-i \epsilon_q t/ R)
\left(
\begin{array}{c}
j_0(x_q\, r/R) \\
i\, \beta_q\, \bm{\sigma}\cdot\hat{\bm{r}}\, j_1(x_q\, r/R)
\end{array}
\right)
\frac{\chi_q}{\sqrt{4 \pi}},
\label{eq:quarkwf}
\end{eqnarray}
with
\begin{eqnarray}
{\cal N}^{-2}_q&=&2 R^3 j_0^2(x_q)[\Omega_q (\Omega_q -1)+R\,m_q^* /2]/x^2_q,\\
\epsilon_q &=& \Omega_q + g^q_\omega\,\omega_0 \, R, \\
\beta_q &=& \sqrt{\frac{\Omega_q - R\, m^*_q}{\Omega_q + R\, m^*_q}}, \\
\Omega_q &=& \sqrt{x^2_q + (R\, m^*_q)^2}, \\
m^*_q &=& m_q - g^q_\sigma\, \sigma,
\end{eqnarray}
where $R$ is the bag radius, $j_0(x)$ and $j_1(x)$ the spherical
Bessel functions, and $\chi_q$ the quark spinor.
The value of
$x_q$ is determined from the boundary condition on the bag surface;
\begin{eqnarray}
j_0(x_q) = \beta_q\, j_1(x_q).
\label{eq:boundcond}
\end{eqnarray}
The energy of a baryon
with ground state quarks is
given by
\begin{eqnarray}
E_B &=& \sum_q  \frac{\Omega_q}{R_B} - \frac{Z_b}{R_B} + \frac{4\pi R_B^3}{3} B_B,
\label{eq:bagery}
\end{eqnarray}
where $B_B$ is the bag constant,
and $Z_B$ is a phenomenological
constant introduced to take into account the zero-point motion of the baryon.
We use the subscript `$B$' to denote the species of a baryon.
The effective mass of a baryon $B$ in medium is given by
\begin{eqnarray}
m^*_B = \sqrt{E^2_B - \sum_q  \left(\frac{x_q}{R_B} \right)^2}.
\label{eq:efmass}
\end{eqnarray}

There are three bag parameters
for each baryon, $R_B$, $B_B$ and $Z_B$.
If one of them can be fixed, the other two can be determined
to reproduce the mass $m_B$ of a baryon $B$ in free space
at a bag radius $R_B$, where $\partial m_B/\partial R_B=0$.
For nucleons in free space, we choose $R_N$ as a free parameter
assuming $R_N \equiv R_p = R_n$.
In actual calculations, we consider a wide range of $R_N$,
$R_N = (0.6,\ 0.8,\ 1.0)$~fm.
For hyperons, we assume the bag radius of hyperons to be the same
as that of nucleons, $R_Y = R_N$,
which then allows us to fix $B_Y$ and $Z_Y$
in the prescribed manner. The bag constant $B_B$ and $Z_B$ 
for $R_0$ =0.6 fm are taken from Ref.~\cite{RHHK} and
those for $R_0$ = 0.8 and 1.0 fm are listed in Table~\ref{tab:BZ}.

\begin{table}[tbp]
\begin{center}
\begin{tabular}{cc|cc|cc|cc} \hline
           &          & $R_0$ = 0.6 fm    &         & $R_0$ = 0.8 fm    &         & $R_0$ = 1.0 fm    &         \\ \hline
~~$B$~~    &$m_B$(MeV)&$B^{1/4}_{B0}$(MeV)&  $Z_B$  &$B^{1/4}_{B0}$(MeV)&  $Z_B$  &$B^{1/4}_{B0}$(MeV)& $Z_B$   \\ \hline
$N$        &   939.0  &~~188.1~~          &~~2.030~~&    ~~157.5~~      &~~1.628~~&   ~~136.3~~       &~~1.153~~\\ \hline
$\Lambda$  &  1115.6  &  197.6            &  1.926  &      164.9        &  1.454  &     142.0         &  0.896  \\ \hline
$\Sigma^+$ &  1189.4  &  202.7            &  1.829  &      168.8        &  1.300  &     145.1         &  0.682  \\ \hline
$\Sigma^0$ &  1192.0  &  202.9            &  1.826  &      168.9        &  1.295  &     145.2         &  0.674  \\ \hline
$\Sigma^-$ &  1197.3  &  203.3            &  1.819  &      169.2        &  1.283  &     145.4         &  0.659  \\ \hline
$\Xi^0   $ &  1314.7  &  207.6            &  1.775  &      172.6        &  1.215  &     147.9         &  0.558  \\ \hline
$\Xi^-   $ &  1321.3  &  207.9            &  1.765  &      172.9        &  1.200  &     148.1         &  0.538  \\ \hline
\end{tabular}
\end{center}
\caption{Bag constants $B_{B}$ and phenomenological constants $Z_B$ for
octet baryons to reproduce the free mass of each baryon 
for $R_0$ = 0.6, 0.8 and 1.0 fm.}
\label{tab:BZ}
\end{table}

The coupling constants for up and down quarks with $\sigma$ and $\omega$
mesons can be determined from nuclear saturation properties
by assuming
$g^u_\sigma = g^d_\sigma$, $g^u_\omega=g^d_\omega$, and
$g^s_\sigma = g^s_\omega =0$.
That is, $g^u_\sigma$ and $g^u_\omega$ can be determined to reproduce
the binding energy per nucleon $E/A = 16$ MeV at 
the nuclear saturation density $\rho_0 = 0.17$ fm$^{-3}$.

In the QMC model, where the bag constant $B_B$ is density-independent,
the nucleon mass at the saturation density is predicted to be larger than
the widely accepted range $m^*_N = (0.7-0.8)\, m_N$
and the compression modulus $K$ is obtained to be
smaller than the empirical values $K = (200 - 300)$ MeV.
On the other hand,
the MQMC model, having density-dependence in the bag constant
with an additional parameter $g'^B_\sigma$, can
produce both the effective mass and the compression modulus
in the reasonable ranges. The density dependent bag constant
can be expressed as \cite{mqmc96}
\begin{eqnarray}
B_B (\sigma) = B_{B}\exp\left(-4\,\sum_{q=u,d}n_q{g'}^B_\sigma\sigma/m_B\right),
\label{eq:mqmcbag}
\end{eqnarray}
where $n_q$ is the number of $u$ and $d$ quarks in a baryon $B$
and $m_B$ is its free mass.
The coupling constants for both the QMC and MQMC models
and the resulting nuclear
matter properties can be found in Table I of Ref.~\cite{theta05}. 


In the subsequent sections, we shall see that the in-medium bag radius
$R_B(\rho)$ plays an important role.
It is defined as the bag radius where
\begin{eqnarray}
\left. \frac{\partial m^*_B}{\partial R_B}\right|_{R_B = R_B(\rho)} = 0.
\label{eq:minimum}
\end{eqnarray}
It is known that there is a sharp contrast between the QMC and the MQMC models 
in the density-dependence of $R_N(\rho)$ \cite{JJ}.
$R_N(\rho)$ from QMC decreases just a little while that from MQMC
increases by about $(10 \sim 20)$~\% at the saturation density~\cite{mqmc96}.
We shall show in Sec.III there are correlations between 
the bag radius in medium $R_B (\rho)$
and the magnetic moments in medium $\mu _B (\rho)$.

\subsection{Magnetic moment of baryons}

The nucleon bags in both QMC and MQMC models become a simple MIT bag 
in free space. We thus briefly describe first the calculation of the
magnetic moments of baryons by using the MIT bag model, 
whose detailed explanation can be found,
for example, in Ref.~\cite{textbook}.

The magnetic moment operator can be written as
\begin{eqnarray}
\sum_i {\hat{\bm \mu}}_i =\sum_i \frac{\hat{Q}_i}{2}{\bm r}_i \times{\bm \alpha},
\label{eq:mag_op}
\end{eqnarray}
where $\hat{Q}_i$ and ${\bm r}_i$ are the charge and the position operators
of the $i$-th quark ($i$ = 1, 2, 3) in the bag, and
${\bm \alpha} = \gamma_0 {\bm \gamma}$.
The normalized SU(6) wave function of a spin-up proton is
given as
\begin{eqnarray}
\left| \Psi_p \right> &=& \frac{1}{3\sqrt 2} \{
2u^\uparrow(1)u^\uparrow(2)d^\downarrow(3)
- u^\uparrow (1)u^\downarrow(2)d^\uparrow(3)
- u^\downarrow (1)u^\uparrow(2)d^\uparrow(3) \nonumber \\
&&
- u^\uparrow(1)d^\uparrow(2)u^\downarrow(3)
+ 2 u^\uparrow(1)d^\downarrow(2)u^\uparrow(3)
- u^\downarrow(1)d^\uparrow(2)u^\uparrow(3) \nonumber  \\
&&
- d^\uparrow(1)u^\uparrow(2)d^\uparrow(3)
- d^\uparrow(1)u^\uparrow(2)d^\uparrow(3)
+ 2d^\downarrow(1)u^\uparrow(2)d^\uparrow(3) \},
\label{eq:wavefunc}
\end{eqnarray}
where $q^{\rm x}(i)$ denotes a quark wave function given
by Eq.~(\ref{eq:quarkwf}) with its spin state
${\rm x}$ ($= \uparrow$ or $\downarrow$).
The magnetic moment of a proton
then reads
\begin{eqnarray}
{\bm \mu}_p &=& \left< \Psi_p \right|
~\sum_i \hat{\bm \mu}_i ~ \left| \Psi_p \right> \nonumber \\
&=&
\frac e2 \int d^3 r~{u^\uparrow}^\dag ~({\bm r}\times{\bm \alpha})~u^\uparrow
\nonumber \\
&\equiv& \frac e2 D_u\,
{\bm e}_z,
\end{eqnarray}
where ${\bm e}_z$ is the unit vector along the $z$-axis and
the integral $D_q$ is given by
\begin{eqnarray}
D_q = \frac 43\, {\cal N}_q^2\, \beta_q
\left( \frac {R_B}{x_q} \right )^4 \int_0^{x_q}
y^3 j_0(y)j_1(y)dy.
\label{eq:Dq}
\end{eqnarray}

Wave functions for other baryons can be obtained by acting
the SU(3) shift operators $\hat{T}_{\pm}$, $\hat{U}_{\pm}$ and $\hat{V}_{\pm}$
successively to the proton wave function \cite{greiner1} as follows:
\begin{eqnarray}
| n \uparrow \rangle       &=& \hat T_- | p \uparrow \rangle, ~~~~~
|\Sigma^+ \uparrow \rangle = \hat U_- | p \uparrow \rangle,
\nonumber \\
|\Sigma^0 \uparrow \rangle &=& \hat T_- | \Sigma^+ \uparrow \rangle, ~~~~~
|\Sigma^- \uparrow \rangle = \hat T_- | \Sigma^0 \uparrow \rangle,
\nonumber \\
|\Xi^- \uparrow \rangle &=& \hat U_- | \Sigma^- \uparrow \rangle, ~~~~~
|\Xi^0 \uparrow \rangle = \hat T_+ | \Xi^- \uparrow \rangle,
\end{eqnarray}
and $|\Lambda \uparrow \rangle$ can be obtained by the orthonormality condition.
Once the wave functions are obtained, the calculation
of the matrix elements is straightforward.
Applying the magnetic moment operator to the wave functions of octet baryons,
we obtain
\begin{eqnarray}
\mu_p &=& \frac e2 D_u, \nonumber \\
\mu_n &=& - \frac e3 D_u,  \nonumber \\
\mu_{\Lambda} &=& - \frac e6 D_s,  \nonumber \\
\mu_{\Sigma^+} &=& \frac e6 \left[\frac 83 D_u +\frac 13 D_s\right], \nonumber \\
\mu_{\Sigma^0} &=& \frac e6 \left[\frac 23 D_u +\frac 13 D_s\right], \nonumber \\
\mu_{\Sigma^-} &=& \frac e6 \left[-\frac 43 D_u + \frac 13 D_s\right], \nonumber \\
\mu_{\Xi^0} &=& -\frac e3 \left[\frac 13 D_u +\frac 23 D_s\right], \nonumber \\
\mu_{\Xi^-} &=& \frac e6 \left[\frac 13 D_u -\frac 43 D_s\right].
\label{eq:mm}
\end{eqnarray}

In nuclear matter ($\rho\neq 0$),
the $\sigma$- and $\omega$-mesons acquire non-vanishing values of their mean fields,
which causes changes in the effective masses of quarks
($m^*_q = m_q - g^q_\sigma\, \sigma$)
as well as in
other quantities such as ${\cal N}_q$, $\beta_q$, $x_q$ and $R_B(\rho)$.
Thus, $D_q$ and the resulting values of the magnetic moments of baryons
depend on the nuclear density.

\section{Results \label{sec:result}}

\begin{table}[tbp]
\begin{center}
\begin{tabular}{ccccccccc}\hline
$R_0$ (fm) &\phantom{cc} $p$\phantom{cc} &\phantom{cc} $n$\phantom{cc} &
\phantom{cc} $\Lambda$\phantom{cc} &\phantom{cc} $\Sigma^+$\phantom{cc} &
\phantom{cc}$\Sigma^0$\phantom{cc} &\phantom{cc} $\Sigma^-$\phantom{cc} &
\phantom{cc} $\Xi^0$\phantom{cc} &\phantom{cc} $\Xi^-$\phantom{cc} \\ \hline
0.6 & 1 & $-0.667$ & $-0.303$ & 0.977 & 0.320 & $-0.337$ & $-0.615$ & $-0.291$ \\
0.8 & 1 & $-0.667$ & $-0.295$ & 0.977 & 0.318 & $-0.341$ & $-0.607$ & $-0.282$ \\
1.0 & 1 & $-0.667$ & $-0.288$ & 0.975 & 0.316 & $-0.344$ & $-0.598$ & $-0.272$ \\
\hline
Exp.& 1 & $-0.685$ & $-0.219$ & 0.880 &  & $-0.415$ & $-0.448$ & $-0.233$ \\
\hline
\end{tabular}
\end{center}
\caption{The ratios $\mu_B/\mu_p$ for octet baryons 
in free space for three choices of $R_0$ values.  
The experimental values of the magnetic moments 
are taken from Ref.~\cite{pdg04}.}
\label{tab:mm_free}
\end{table}

Before we present the results for $\mu_B (\rho)$ in medium, let us first show 
the values of $\mu_B (\rho=0)$ in free space calculated by using 
Eq.~(\ref{eq:mm}).
The ratios of $\mu_B/\mu_p$ in free space are listed 
in Table~\ref{tab:mm_free}, where $R_0$ denotes the bag radius
in free space; $R_0 = R_B (0)$.
Due to the difference in the values of parameters,
the magnetic moments of this work are slightly different from those given in
Ref.~\cite{textbook}.

Let us introduce a quantity $r_B (\rho)$ defined as the ratio of 
the magnetic moment of a baryon $B$
in medium of density $\rho$ relative to its free space value,
\begin{equation}
r_B(\rho) \equiv \frac{\mu_B(\rho)}{\mu_B}.
\end{equation}
We list the values of $r_B (\rho_0 )$'s at the saturation density $\rho_0$
obtained from the QMC and the MQMC models in Table~\ref{tab:mm_qmc}
and Table~\ref{tab:mm_mqmc}, respectively.
To check the consistence of the results against different choices 
of the bag radius,
we show the ratios $r_B(\rho_0)$ for different values of $R_0$.
The $R_0$-dependence of $r_B(\rho_0)$ in the QMC model is found to be
rather small (less than 5 \%), while that in the MQMC model is even smaller
(less than 2 \%). Thus in the forthcoming discussion we use $R_0 = 0.8$ fm.
\begin{table}[tbp]
\begin{center}
\begin{tabular}{cccccccc}\hline
$R_0$ (fm) &\phantom{cc} $r_N$ \phantom{cc} &
\phantom{cc} $r_\Lambda$\phantom{cc} &\phantom{cc} $r_{\Sigma^+}$\phantom{cc} &
\phantom{cc} $r_{\Sigma^0}$\phantom{cc} &\phantom{cc} $r_{\Sigma^-}$\phantom{cc} &
\phantom{cc}$r_{\Xi^0}$\phantom{cc} &\phantom{cc} $r_{\Xi^-}$\phantom{cc} \\ \hline
0.6 & 1.029 & 0.993 & 1.037 & 1.027 & 1.057 & 1.015 & 0.980 \\
0.8 & 1.053 & 0.997 & 1.053 & 1.040 & 1.077 & 1.021 & 0.976 \\
1.0 & 1.071 & 0.999 & 1.067 & 1.051 & 1.095 & 1.027 & 0.970 \\ \hline
\end{tabular}
\end{center}
\caption{The ratios $r_B(\rho_0) = \mu_B(\rho_0)/\mu_B$,
where $\mu_B(\rho_0)$ and $\mu_B$ are the magnetic moments at normal
nuclear matter density $\rho_0$ and in free space, respectively,
are tabulated for three $R_0$ values.
$\mu_B(\rho_0)$ is calculated by the QMC model.}
\label{tab:mm_qmc}
\end{table}

\begin{figure}[tbp]
\begin{center}
\epsfig{file=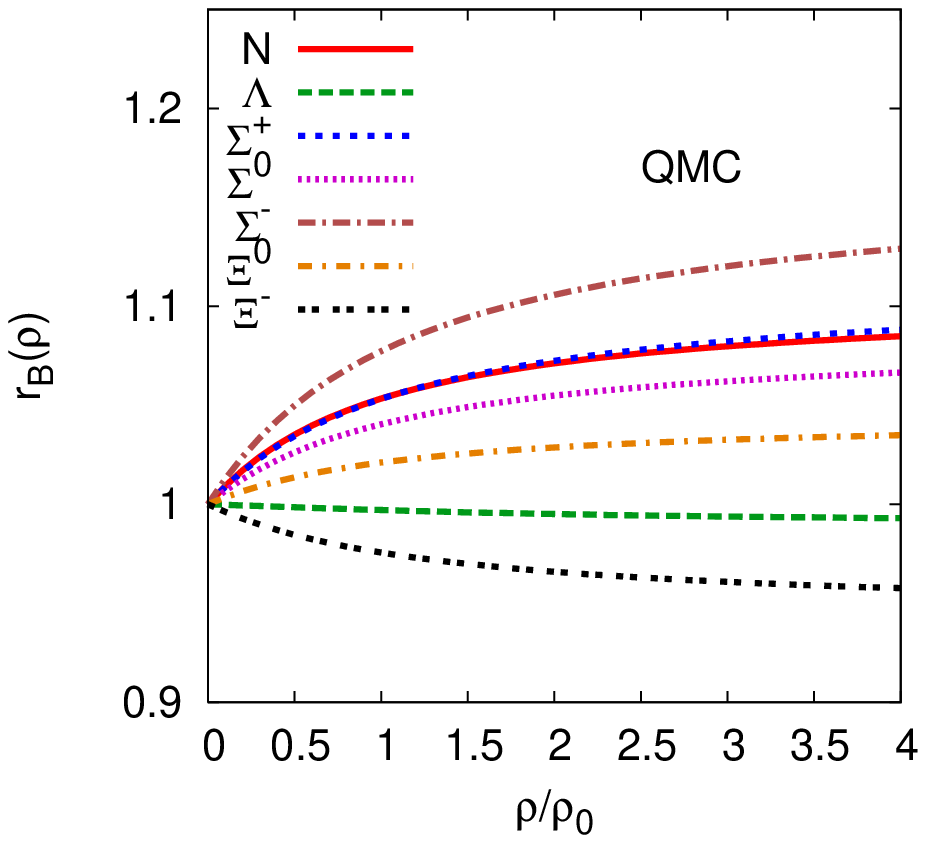, width=12cm}
\epsfig{file=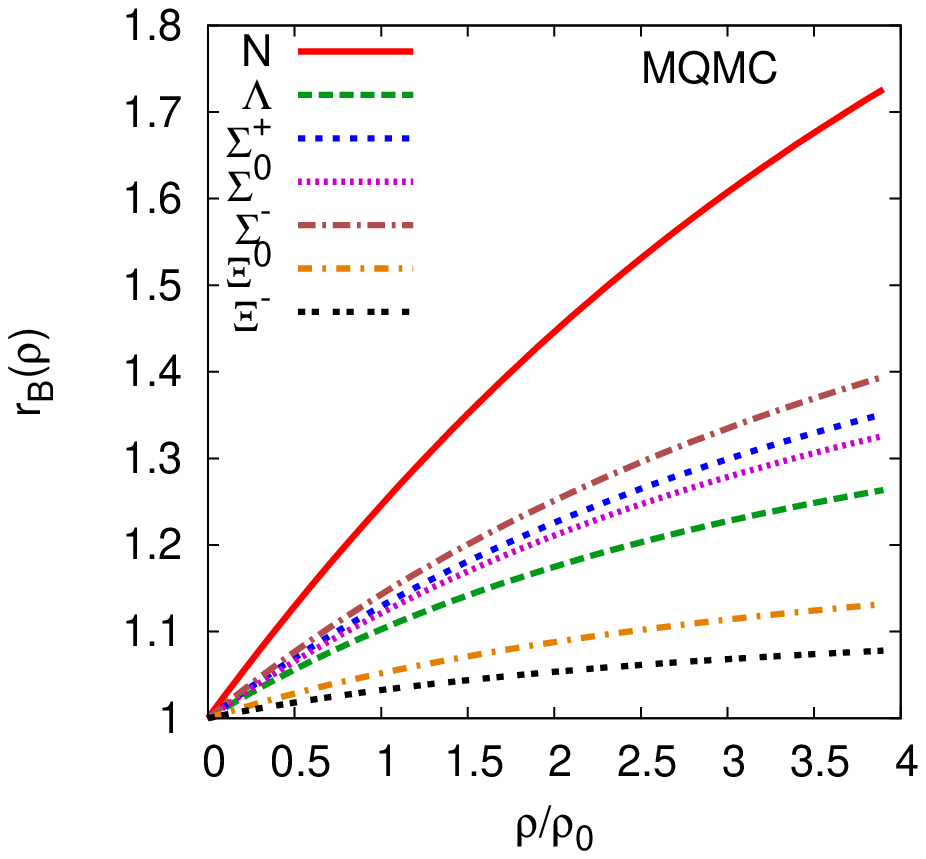, width=12cm}
\end{center}
\caption{The ratios $r_B (\rho) = \mu_B (\rho) / \mu_B$ of the
magnetic moments $\mu_B (\rho)$ of octet baryons in dense matter
calculated by the QMC and MQMC models are plotted as a function of density
in the upper and lower panels, respectively.}
\label{fig:mm_result}
\end{figure}

As seen in Table ~\ref{tab:mm_qmc}, the magnetic moments of baryons 
at the saturation density in the QMC model change only by
a few percents from those in free space for all the baryons.
This behavior agrees with the results of Ref. \cite{st95}.
In particular, the magnetic moment of $\Lambda$ in matter 
remains almost unchanged from that in free space
with only about $(0.1 \sim 0.7)$~\% decrease
at the saturation density. Even if we increase the matter density
up to 4 times the saturation density, the change is found to be
less than 2~\% for $\Lambda$ as seen in the upper panel of Fig.~1. 
The reason is not difficult to understand.
As the $s$-quark does not couple to the $\sigma$ and $\omega$ mesons,
the effective mass of the s-quark remains unchanged, $m_s^*=m_s$.
Thus, as can be seen from Eq.~(\ref{eq:Dq}),
the only medium effect on $\mu_\Lambda$ is through the change in the bag radius.
The bag radius $R_B$ in the QMC model, however, decreases
only slightly as density increases.
In our calculation, the bag radius of a $\Lambda$ hyperon at the saturation
density is about 99.6~\% of that in free space.
As a result, $D_s$ remains almost constant with respect to $\rho$,
and consequently the density-dependence of $\mu_B$
coming from that of $D_s$
is very small as seen in Table~\ref{tab:mm_qmc}.
In the upper panel of Fig.~\ref{fig:mm_result},
we show the density dependence of the magnetic moments
of octet baryons (with $R_0 = 0.8$ fm) up to $\rho=4\ \rho_0$.
We find that the density dependence is rather small even at $\rho=4\ \rho_0$
with a change of about 10\ \%.

One can notice from Fig.~\ref{fig:mm_result} that only $r_B (\rho)$'s for
$B= \Lambda$ and $\Xi^-$ are smaller than unity. This can be understood
as follows. Since $D_u$ ($D_s$) is simply proportional to 
$\mu_p$ ($\mu_{\Lambda}$), one can envision $D_u$ and $D_s$ 
from the curves for $r_N (\rho)$ and $r_{\Lambda} (\rho)$,
respectively, shown in Fig.~\ref{fig:mm_result}. 
$D_u (\rho)$ increases with density as $r_N (\rho)$ does, but 
$D_s (\rho)$ remains almost constant with respect to density 
as $r_{\Lambda}(\rho)$ does in the case of the QMC model.
Equation (\ref{eq:mm}) shows that $\mu_n$, $\mu_{\Lambda}$,
$\mu_{\Sigma^-}$, $\mu_{\Xi^0}$, and $\mu_{\Xi^-}$ are negative-valued.
Among these, only $\mu_{\Lambda}$ and $\mu_{\Xi^-}$ decrease 
in magnitude with density (in QMC model),
which causes the ratio $r_B (\rho)$ becoming smaller than unity 
for $\Lambda$ and $\Xi^-$.


\begin{table}[tbp]
\begin{center}
\begin{tabular}{cccccccc}\hline
$R_0$ (fm) &\phantom{cc} $r_N$ \phantom{cc} &
\phantom{cc} $r_\Lambda$\phantom{cc} &\phantom{cc} $r_{\Sigma^+}$\phantom{cc} &
\phantom{cc} $r_{\Sigma^0}$\phantom{cc} &\phantom{cc} $r_{\Sigma^-}$\phantom{cc} &
\phantom{cc}$r_{\Xi^0}$\phantom{cc} &\phantom{cc} $r_{\Xi^-}$\phantom{cc} \\ \hline
0.6 & 1.237 & 1.106 & 1.125 & 1.118 & 1.135 & 1.051 & 1.037 \\
0.8 & 1.246 & 1.103 & 1.129 & 1.121 & 1.143 & 1.052 & 1.032 \\
1.0 & 1.254 & 1.099 & 1.134 & 1.124 & 1.151 & 1.053 & 1.027 \\ \hline
\end{tabular}
\end{center}
\caption{The ratios $r_B(\rho_0)=\mu_B(\rho_0)/\mu_B$ 
calculated with the MQMC model are tabulated for three different 
values of $R_0$, where $\mu_B(\rho_0)$ and $\mu_B$ are the magnetic moments 
at the normal nuclear matter density and in free space, respectively.}
\label{tab:mm_mqmc}
\end{table}

As plotted in the lower panel of Fig.~\ref{fig:mm_result},
the magnetic moments calculated by the MQMC model are
quite different from those in the QMC model. 
The magnetic moments of all the baryons increase uniformly, but
the slope of $r_N(\rho)$ is steeper than that of other $r_B(\rho)$'s.
The magnetic moments at $\rho=\rho_0$ are listed in Table ~\ref{tab:mm_mqmc}.
At the saturation density, the magnetic moment of a nucleon in the MQMC model
increases by about 25~\% from its free space value,
which is much larger than the increase observed
with the QMC model ($3 \sim 7$ \%).
Also, the MQMC models predicts the value of $\mu_\Lambda$ 
at the saturation density to increase by about 10~\%,
which is in sharp contrast with the QMC prediction,
in which case $\mu_\Lambda$ in medium remains almost the same as 
the free space value.

Comparing Table \ref{tab:mm_qmc} and Table \ref{tab:mm_mqmc} shows
the values of $r_\Sigma$'s and $r_\Xi$'s are
considerably different depending on the model used.
The differences in the $r_B (\rho )$ values calculated 
by the QMC and the MQMC models can be attributed to the fact that 
the bag radii change considerably in the MQMC model.
Since the bag constant decreases very rapidly with density in the MQMC model, 
the bag radius increases with density to satisfy
the minimization condition, $\partial m^*_B/ \partial R = 0$.

\begin{figure}[tbp]
\begin{center}
\epsfig{file=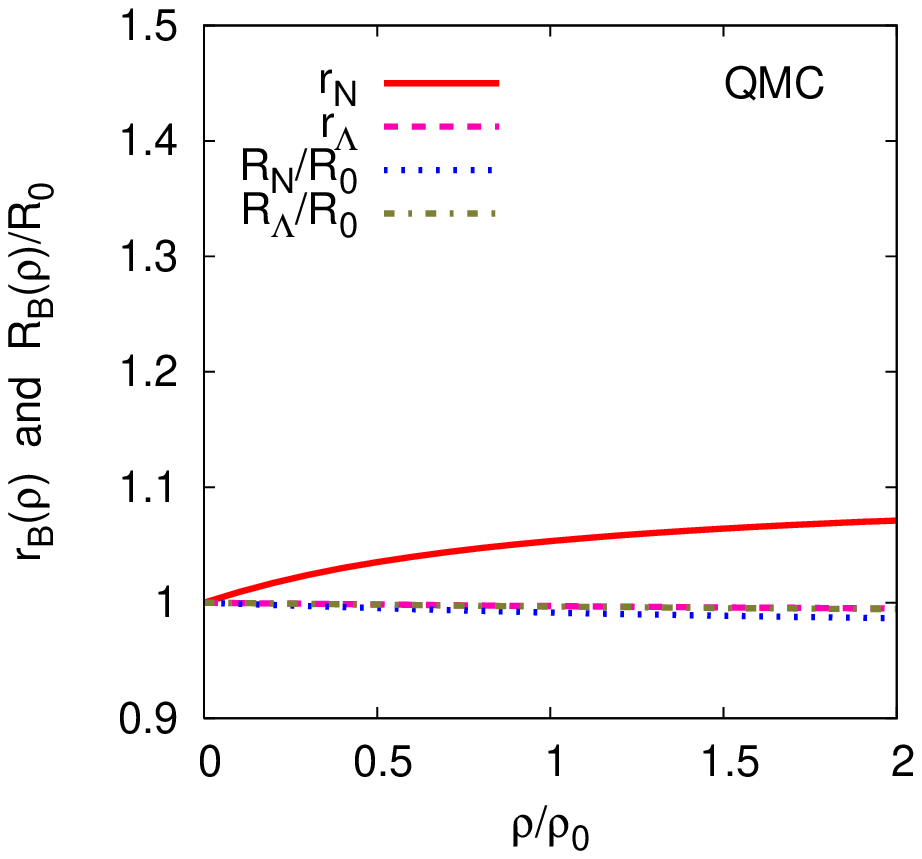, width=12cm}
\epsfig{file=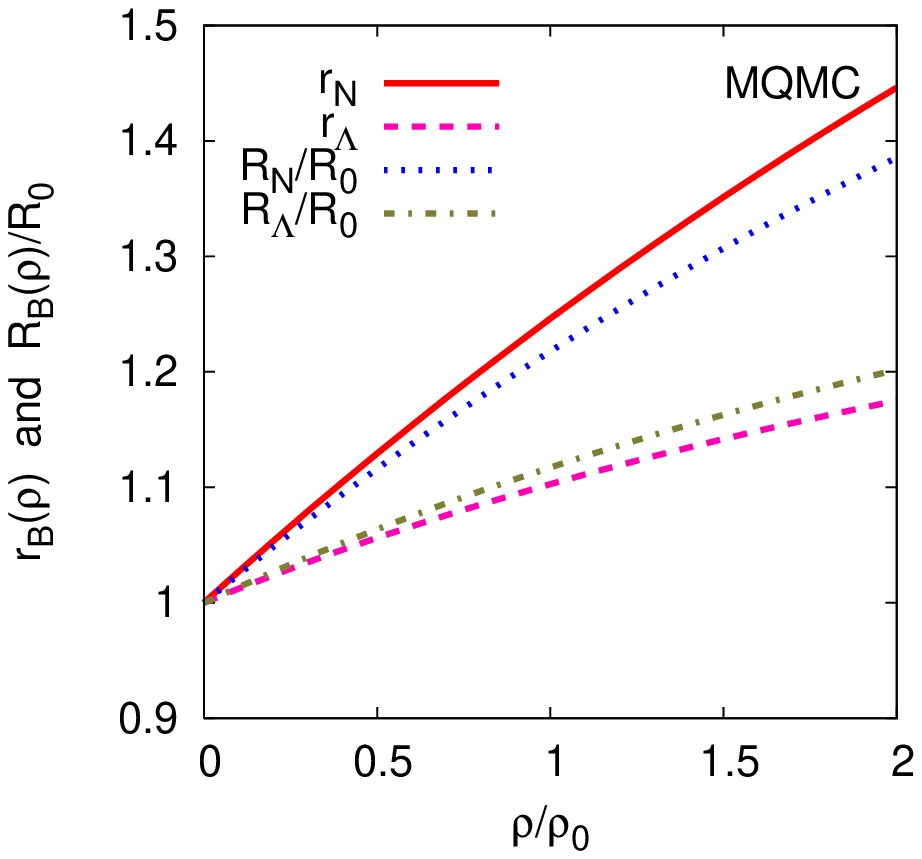, width=12cm}
\end{center}
\caption{Comparison of $R_B (\rho )/R_0$ and $r_B (\rho)$ ($B=N$ and $\Lambda$)
as a function of density. $R_0$ is chosen as $0.8$ fm. In the upper panel 
the dashed curve overlaps with the dash-dotted curve.}
\label{fig:RM_compare}
\end{figure}

To see how much the magnetic moments are correlated with
the values of the bag radius $R_B (\rho )$, 
we plot in Fig. \ref{fig:RM_compare}
$r_B(\rho)=\mu_B(\rho)/\mu_B$ for $B=N$ and $\Lambda$ together with
the ratio $R_B(\rho)/R_0$.
(We consider here only two cases of $B=N$ and $\Lambda$ because 
$\mu_N (\rho)$ and $\mu_{\Lambda} (\rho)$ are simply proportional 
to $D_u$ and $D_s$, respectively.)
The results from the QMC and MQMC models are plotted 
in the upper and lower panels, respectively.
The dotted curve in the upper panel of Fig. \ref{fig:RM_compare}
represents $R_N (\rho )/R_0$ calculated by QMC.
We can see that $R_N (\rho )/R_0$ remains almost the same as the value
in free space. As a result, $r_N(\rho)$ denoted by the solid curve
also changes just a little, less than 10\%.
For a $\Lambda$ hyperon, the dashed curve denoting the ratio
$r_\Lambda(\rho) = \mu_\Lambda(\rho) / \mu_\Lambda$
overlaps with the dash-dotted curve for $R_{\Lambda}/R_0$, and so
they are not distinguishable in Fig. \ref{fig:RM_compare}. 
Since $R_{\Lambda}/R_0$ remains almost constant with respect to $\rho$,
$r_{\Lambda}(\rho )$ also remains almost constant.
The lower panel of Fig. \ref{fig:RM_compare} also shows that
the bag radii and the magnetic moments of both $N$ and $\Lambda$ 
in the MQMC model are strongly correlated.
The behaviour of $r_N$ ($r_{\Lambda}$) is rather similar to that of
$R_N / R_0$ ($R_{\Lambda}/R_0$).
It is known that for massless quarks
$D_q$ is proportional to the bag radius $R$,
$D_q \propto R$ \cite{textbook}.
For general cases where $m_q^*\neq 0$, the analysis becomes complicated.
Figure \ref{fig:RM_compare}, however, indicates that
such a relation still remains valid to some extent
for the cases considered in this work.

\section{Summary}

We have considered in this work the changes of the magnetic moments 
of octet baryons in dense matter for the first time in the 
framework of the quark-meson coupling model.
Both octet baryons and nuclear matter are treated in a consistent manner
by using the quark-meson coupling models.
Model dependence is investigated by employing the QMC and the MQMC models,
which differ in the density dependence of the bag constant.
The QMC model predicts that the magnetic moments of octet baryons
at the saturation density change only by a few percents 
from those in free space.
In particular, the magnetic moment of $\Lambda$ practically
does not change from the free-space value in the QMC model.
On the other hand, we obtain quite different results from the MQMC model.
In the MQMC model, the magnetic moment of a nucleon
at the saturation density
increases by about 25~\% from the free-space value,
and the magnetic moment of $\Lambda$ also
increases by about 10~\%.
A similar amount of increase in the magnetic moment is also observed
for other hyperons.
The reason for this model dependence can be ascribed mainly to the behavior of
the bag radius in nuclear matter, which comes from the change in 
the bag constant with respect to the density.
The self-consistency equations in the QMC and MQMC models are highly nonlinear.
Thus it is not straightforward to see 
how magnetic moments are related to the bag radius.
However, by comparing the density dependence of the magnetic moment 
with that of the bag radius for each baryon, 
we find that the two quantities behave rather similarly as functions of density.
In the QMC model, the bag radii of the baryons decrease only 
slightly at the saturation density from the free-space values, by about 1~\%.
The magnetic moments calculated with the QMC model change
more or less in the same ratios.
In the MQMC model, the bag radii and the magnetic moments behave very similarly
and increase by about 10 -- 20~\% at the nuclear saturation density.
Our present results for magnetic moments in nuclear matter may not be
directly comparable to the magnetic moments of baryons in finite nuclei. 
However, our results show that there can be significant changes of the magnetic
moments in nuclei, and further investigations are needed to
to understand the medium modification of the EM properties of baryons.
In particular, it is desirable to measure the magnetic moments of 
these octet baryons in nuclei, which can also
give us information on various medium effects on hyperons \cite{saito97, lu98}. 


\section*{Acknowledgments}
This work was supported in part by the Korea Science and Engineering Foundation 
grant funded by the Korean Government (MOST) (No. M20608520001-07B0852-00110).
CYR was supported by the Korea Research Foundation Grant funded
by the Korean Government (MOEHRD) (KRF-2006-214-C00015).
CHH was supported by the Korea Research Foundation Grant funded 
by the Korean Government (MOEHRD) (KRF-2006-312-C00506).
TSP was supported in part by KOSEF Basic Research Program 
with the grant No. R01-2006-10912-0.


\begin{thebibliography}{100}
\bibitem{emc83} J.~J. Aubert {\it et al.}, Phys. Lett. {\bf 123B} (1983) 275. 
\bibitem{mulders90} P.~J. Mulders, Phys. Reports {\bf 185} (1990) 83.
\bibitem{jlab07} S. Strauch,
arXiv:0709.4034v1 [nucl-ex].
\bibitem{cheon92} Il-T. Cheon and M.~T. Jeong, 
J. Phy. Soc. Jpn. {\bf 61} (1992) 2726. 
\bibitem{frank96} M.~R. Frank, B.~K. Jennings and G.~A. Miller, 
Phys. Rev. C {\bf 54} (1996) 920.
\bibitem{lu99} D.~H. Lu, K. Tsushima, A.~W. Thomas, A.~G. Williams and
K. Saito, Phys. Rev. C {\bf 60} (1999) 068201.
\bibitem{yak03} U.~T. Yakhshiev, U.-G. Mei\ss ner, A. Wirzba, Eur. Phys.
J. {\bf A 16} (2003) 569.
\bibitem{smith04} J.~R. Smith and G.~A. Miller, 
Phys. Rev. C {\bf 70} (2004) 065205.
\bibitem{horikawa05} T. Horikawa and W. Bentz, 
Nucl. Phys. {\bf A 762} (2005) 102.
\bibitem{kc03} K.~S. Kim and M.~K. Cheoun, 
Phys. Rev. C {\bf 67} (2003) 034603.
\bibitem{kw03} K.~S. Kim and L.~E. Wright, 
Phys. Rev. C {\bf 68} (2003) 027601.
\bibitem{qmc88} P.~A.~M. Guichon, 
Phys. Lett. {\bf B 200} (1988) 235.
\bibitem{mqmc96} X. Jin and B.~K. Jennings, 
Phys. Rev. C {\bf 54} (1996) 1427.
\bibitem{ST-PLB} K. Saito and A. W. Thomas, 
Phys. Lett. {\bf 327} (1994) 9.
\bibitem{MPP} D. P. Menezes, P. K. Panda, C. Providencia, 
Phys. Rev. C {\bf 72} (2005) 035802.
\bibitem{RHHJ} C. Y. Ryu, C. H. Hyun, S. W. Hong, and B. K. Jennings, 
Eur. Phys. J. A {\bf 24} (2005) 149.
\bibitem{st95} K. Saito and A.~W. Thomas, 
Phys. Rev. C {\bf 51} (1995) 2757.
\bibitem{JJ} X. Jin and B.~K. Jennings, 
Phys. Lett {\bf B 374} (1996) 13.  
\bibitem{tamura07} H. Tamura, invited talk at international conference
{\it ``Chiral07"}, Osaka, Japan, Nov. 13--16, 2007.
\bibitem{RHHK} C.~Y. Ryu, C.~H. Hyun, S.~W. Hong, and B.~T. Kim,
Phys. Rev. C {\bf 75} (2007) 055804.
\bibitem{theta05} C.~Y. Ryu, C.~H. Hyun, J.~Y. Lee and S.~W. Hong,
Phys. Rev. C {\bf 72} (2005) 045206.
\bibitem{textbook} W. Greiner, S. Schramm and E. Stein,
{\it Quantum Chromodynamics}, 3rd ed. (Springer-Verlag, New York, 2007).
\bibitem{greiner1} W. Greiner and B. Muller, {\it Quantum mechanics:
Symmetries} (Springer-Verlag, New York, 1994).
\bibitem{pdg04} Particle Data Group, S. Eidelman et al.,
Phys. Lett. B {\bf 592} (2004) 1.
\bibitem{saito97} K. Saito, M. Oka and T. Suzuki,
Nucl. Phys A {\bf 625} (1997) 95.
\bibitem{lu98} D.~H. Lu, K. Tsushima, A.~W. Thomas, A.~G. Williams
and K. Saito, Nucl. Phys. A {\bf 634} (1998) 443.
\end{thebibliography}
\end{document}